\documentclass[
    reprint,
    preprintnumbers,
    superscriptaddress,
    nofootinbib,
     amsmath,amssymb,
     aps,
     prd,
    floatfix,
    longbibliography]{revtex4-2}
    
    \usepackage{graphicx}
    \usepackage[caption=false]{subfig}
    \usepackage[usenames,dvipsnames]{xcolor} 
    \usepackage{pgfplots}
    \usepackage[utf8]{inputenc}
    \usepackage{color}
    \usepackage{hyperref}
    \usepackage[normalem]{ulem} 
    \usepackage{physics}
    \usepackage{enumitem}
    \usepackage{comment}
    \usepackage{bm}
    \usepackage{aas_macros}
    \usepackage[capitalise]{cleveref}
    \usepackage{tensor}
    \usepackage{appendix}
    \usepackage{orcidlink}
\hypersetup{
  breaklinks = true,
  colorlinks   = true, 
  urlcolor     = blue, 
  linkcolor    = blue, 
  citecolor   = blue 
}
    \allowdisplaybreaks[1]

\usepackage{float}

\newcommand{\splitatcommas}[1]{%
  \begingroup
  \begingroup\lccode`~=`, \lowercase{\endgroup
    \edef~{\mathchar\the\mathcode`, \penalty0 \noexpand\hspace{0pt plus 1em}}%
  }\mathcode`,="8000 #1%
  \endgroup
}

\begin{document}

\title{Coupled quintessence from an axion dark sector}

\author{Rayff de Souza\,\orcidlink{0000-0003-3526-8763}}
\email{rayffsouza@on.br}
\affiliation{Observatório Nacional, Rio de Janeiro, RJ, 20921-400, Brazil}
\affiliation{School of Physics and Astronomy, University of Nottingham, Nottingham, NG7 2RD, United Kingdom}

\author{Edmund J. Copeland\, \orcidlink{0000-0003-3959-6051}}
\email{ed.copeland@nottingham.ac.uk}
\affiliation{School of Physics and Astronomy, University of Nottingham, Nottingham, NG7 2RD, United Kingdom}

\author{Jailson Alcaniz\,\orcidlink{0000-0003-2441-1413}}
\email{alcaniz@on.br}
\affiliation{Observatório Nacional, Rio de Janeiro, RJ, 20921-400, Brazil}

\date{\today}

\begin{abstract}
Recent observational data arising from the DESI collaboration \cite{DESI:2025zgx} has hinted at a possible departure from the standard $\Lambda$CDM cosmological model, preferring instead the presence of a dynamical dark energy component. Specifically, the associated equation of state of the dark energy features a crossing into the so-called phantom regime, which is challenging to accommodate in canonical single scalar-field scenarios. However, this behavior can be effectively described by an interacting dark sector, where the specific dark energy equation of state remains above the phantom divide whilst the dark matter component deviates from the standard cold dark matter evolution. In this work, we explore this possibility in the context of an axion dark sector, where both the dark energy and dark matter are represented by two interacting axion-like fields. We show that given the required mass hierarchy for these fields to play such roles, their dynamics can be effectively placed in the coupled quintessence framework, where their motion follows from a sourced continuity equation in the fluid description. In this regime, we perform a statistical analysis of this scenario with current data, finding that a sub-Planckian dark energy axion decay constant stays well within the observational bounds without the need to fine-tune the associated field's initial conditions. We also perform a comparison with $\Lambda$CDM, where we find that the model provides a better fit to the data while staying competitive from a Bayesian perspective.

\end{abstract}

\maketitle

\section{Introduction}\label{sec:1}
 Recent measurements of Baryon Acoustic Oscillations (BAO) by the DESI Collaboration, in combination with Cosmic Microwave Background (CMB) and Type IA supernovae (SNeIa) data, have challenged the standard $\Lambda$-Cold Dark Matter ($\Lambda$CDM) cosmological model, suggesting their may instead be evidence for a dynamical dark energy component in the Universe \cite{DESI:2025zgx}. Such claims are mostly based on a $\gtrsim 3\sigma$ discrepancy between the time independent evolution of the $\Lambda$CDM equation of state, (EoS), and the mean distribution of the so called CPL \cite{Chevallier:2000qy,Linder:2002et} dynamical dark energy parameterization, $w(a) = w_0 + w_a(1-a)$, where the dark energy EoS parameter, $w(a)$ is the time-dependent ratio between the dark energy pressure and energy density, $w_0$ and $w_a$ are constants, with $a$ being the cosmological scale factor.

In addition, the current favored scenario for CPL features a crossing of the EoS to the so-called phantom regime, i.e $w(z) < -1$ $\forall$ $z > z_c$, with $z_c \simeq 0.5$, a feature which is not reproduced by the simplest  quintessence like single-field scalar field models which are bounded by $-1 \leq w(z) \leq 1$, (recall the redshift $z$ is related to $a$ by $1+z = 1/a$). Consequently, the dark energy density, $\rho_{\rm{DE}}$, defined as $\mathrm{d}\ln\rho_{\rm{DE}} = - 3[1+w(a)]\mathrm{d} \ln a$, reaches a maximum at the phantom-crossing redshift $z_c$ and decreases into the past. Therefore, for the same present-day density $\rho_{\rm{DE,0}}$, the dark energy density in CPL is higher than that of a cosmological constant at very low-redshifts, and becomes smaller for $z \gtrsim  z_c$.

However, background observables -- such as BAO distances or SNIa distance-ladder measurements -- that constrain the cosmic expansion rate $H(z)$, depend only on the overall sum of the components' energy densities. Thus, in order to achieve the $H(z)$ behavior similar to the preferred profile implied by DESI, one approach is to allow for a non-standard evolution of the matter components, by absorbing the necessary dip in the dark energy density (which leads to the phantom-crossing feature) in the energy evolution of the matter fields -- see \cite{Mishra:2026tzn,Wolf:2025jed} for alternative interpretations. This is the so-called \textit{phantom-mirage} effect \cite{Das:2005yj, Caldwell:2025inn,Liu:2025bss}, which allows us to interpret current data in a beyond-$\Lambda$CDM framework without invoking a phantom dark energy component.

A simple way to achieve a non-standard matter evolution is to consider non-gravitational interactions of cold dark matter (CDM), such that the cosmological evolution of it's energy density deviates from the exact $\sim a^{-3}$ scaling. Although CDM interactions with the standard model are severely constrained \cite{Kopp:2018zxp, Ilic:2020onu,BuenAbad2022,GluscevicBoddy2018,Boddy2018}, a coupling of CDM to the dark energy component remains an intriguing possibility and has received much attention in light of recent observational results \cite{Li2026,Shah2025,Silva2025,Chakraborty2025,You2025,Petri2026} -- see also \cite{SevillanoMunoz:2026jgk} for a recent approach which also leads to a period of early dark energy. In essence, the result of a coupling of CDM to scalar-field dark energy at low-redshifts -- where DE is dynamically relevant -- allows for the CDM energy density to be smaller than the predicted $\Lambda$CDM scaling, for the same present-day DM abundance, which helps to bring the late-time $H(z)$ evolution closer to the preferred DESI behavior. 

These scenarios are generically called coupled quintessence \cite{Amendola:1999er,Pourtsidou:2013nha}, and they are often specified by a fermionic DM field $\psi$ interacting with scalar-field dark energy $\phi$ via a $\phi$-dependent mass term, such that the interaction lagrangian is written as
\begin{equation}\label{eq:1}
    \mathcal{L}_{\mathrm{int}} = m_\psi(\phi)\bar{\psi}\psi\; .
\end{equation}

However, from the model-building perspective, this type of coupling possesses some hierarchy difficulties, mainly regarding these components' mass scales. For DE to be a scalar-field, it has to be extremely light, with $m_\phi \sim H_0 \sim 10^{-33}$ eV, for it to remain frozen for most of cosmic expansion and not run down its potential before it can become the dominant component in the late universe. On the other hand, the standard picture for particle dark matter production in the early universe involves the thermal freeze-out of heavy particles, with masses at least of the order of $m_\psi \gtrsim 10^{3}$ eV in the fermionic case. 

In general, a Yukawa-like interaction such as Eq.~\eqref{eq:1} with $m_\psi(\phi) = m_\psi(1+g \phi/M_P)$ will induce loop-corrections to the DE scalar mass. If these corrections are not controlled, they might spoil the flatness of the DE effective potential, which would be detrimental to its required behavior as the DE component \cite{DAmico:2016jbm,DavidMarsh:2018etu}. In fact, these corrections scale quadratically with $m_\psi$ and, in the simple Yukawa picture, one would obtain \cite{DAmico:2016jbm}
\begin{equation}
    \frac{\Delta m_\phi}{m_\phi} \sim g \frac{m_\psi^2}{M_P m_\phi}\sqrt{\ln\left(\frac{\mathcal{M}}{m_\psi}\right)}\;,
\end{equation}
where $\mathcal{M}$ is the renormalization scale. Given the required hierarchies of DE and DM masses, this ratio can be of order $\sim 10^{23}$ for a DM particle at the GeV scale, unless the coupling constant $g$ is tuned to an extremely small number, which would then make the coupled quintessence scenario unfeasible. In this context, a clean way to forbid large mass corrections to the light DE field is to invoke an underlying symmetry in a low-energy EFT.

A solution to this problem can be introduced in the context of axion monodromies \cite{Kim:2004rp,Silverstein:2008sg,Kaloper:2011jz,DAmico:2016jbm}, which can arise in general compactifications of string theory. In this case, there is a natural mixing between two axions with distinctive mass scales, which can play the role of DE and DM. As it turns out, the axions' discrete shift symmetry protects their masses from large radiative corrections, making them a viable model for a coupled dark sector from a QFT perspective.

In fact, axions as both DE and DM candidates have been extensively proposed in the literature over the last few decades \cite{Freese:1990rb, Frieman:1995pm, Kim:1998vf, Kim:2002zg}. From the DE point of view, a common challenge in these models is that its observational viability is often tied with either a fine-tuning of the axions initial conditions, e.g the field starts its evolution close to the potential maximum \cite{Lin:2025gne}, or the imposition of a super-Planckian decay constant, which  would be troublesome in light of the weak gravity conjecture \cite{Banks:2003sx,Arkani-Hamed:2006emk}. Also, a general feature of an axion DM candidate is that it undergoes fast oscillations at the bottom of its potential when its mass becomes larger than the Hubble rate. From the numerical perspective, this presents challenges for the evolution of the field's dynamics when its mass is higher than $\gtrsim 10^{-27}$ eV, where the background system becomes significantly stiff \cite{Marsh:2011gr,Hlozek:2014lca,UrenaLopez:2015gur, Gaughan:2026xrv, Passaglia:2022bcr}.

As we will show in the following sections, a two-axion coupled dark sector can provide a way out for both of these difficulties. First, if we take the DM axion to be heavy enough, one can average out its fast oscillations to express its dynamics as a coupled dust-like fluid throughout the cosmic range of interest. In turn, its coupling to the light DE axion gives it an effective potential that changes minima from early- to late-times, rendering a sub-Planckian decay constant without the need to fine-tune its initial field value. This latter effect was also recently achieved in the context of axion quintessence via a coupling to dark baryons \cite{Khoury:2025txd, Khoury:2026svx} -- see also \cite{Delaunay:2026fse} for the case of a  coupling to WIMP dark matter.

The rest of this paper is organized as follows: in Sec. \ref{sec:2} we discuss the underlying axion dark sector dynamics as well as the treatment of the fast DM oscillations via an WKB approximation, and show how this model can be described in the coupled quintessence framework between two interacting fluids. In Sec.~\ref{sec: Background evolution and apparent phantom-crossing} we look in more detail at the background evolution of our coupled axion system and show how they can lead to an apparent phantom crossing in the equation of state of the effective dark energy. In Sec. \ref{sec:3} we outline the methodology and datasets used to constrain this coupled axion quintessence model and discuss our results in Sec. \ref{sec:4}. We end with a discussion and final remarks in Sec. \ref{sec:5}.

\section{Coupled axion dark sector}\label{sec:2}

We start off by considering a cosmological dark sector consisting of two interacting axion-like fields. Individually, each axion is subject to their respective instanton potentials \cite{Peccei:1977hh,Weinberg:1977ma,Wilczek:1977pj}, while their interaction is mediated by a potential term that respects the fields' discrete shift symmetry. Hence, we write the full dark sector potential for $\phi$ (DE) and $\chi$ (DM) as
\begin{equation}\label{two-axion potential}
    \begin{split}
        V_\mathrm{DS}(\phi,\chi) & = V_\phi(\phi) + V_\chi(\chi) + V_\mathrm{int}(\phi,\chi)\;,\\
        V_\phi(\phi) &= \Lambda^4_\phi\left[1 - \cos\left( \frac{\phi}{f_\phi} \right) \right]\;,\\
        V_\chi(\chi) &= \Lambda^4_\chi\left[1 - \cos\left( \frac{\chi}{f_\chi} \right) \right]\;,\\
        V_\mathrm{int}(\phi,\chi) &= - \Lambda_{\phi,\chi}^4 \cos\left(\frac{\phi}{f_\phi} - \frac{\chi}{f_\chi}\right)\;,
    \end{split}
\end{equation}
where $f_i$ is each field's decay constant and $\Lambda_i^4 = m_i^2 f_i^2$ with $m_i$ being each field's mass, for $i = \phi, \chi$ repsectively. The potential Eq.~\eqref{two-axion potential} arises from string compactifications and axion monodromy models, where interacting axions with different masses and energy scales are expected \cite{DAmico:2016jbm}. For the case of homogeneous background fields, we define the associated energy densities by 
\begin{equation}\label{def:rhophi}
    \rho_\phi = \frac{\dot{\phi}^2}{2} + \Lambda^4_\phi\left[1 - \cos\left( \frac{\phi}{f_\phi} \right) \right]
\end{equation}
\begin{equation}\label{def:rhochi}
    \begin{split}
    \rho_\chi &= \frac{\dot{\chi}^2}{2} + 
    \Lambda^4_\chi\left[1 - \cos\left( \frac{\chi}{f_\chi} \right) \right] \\
   & - \Lambda_{\phi,\chi}^4 \cos\left(\frac{\phi}{f_\phi} - \frac{\chi}{f_\chi}\right)\;.
    \end{split}
\end{equation}
There is clearly a degree of ambiguity here on who to associate the interaction term with. We have made our decision based on the fact that we wish to highlight the impact of the evolving dark energy field $\phi$ on the dark matter field $\chi$. 

We are interested in investigating whether the $\phi$ and $\chi$ fields can play the role of DE and DM, respectively. To this end, there are some important hierarchies that need to be satisfied. Most notably, the DE axion needs to be light enough so that it does not roll all the way down its cosine potential at some point during cosmic evolution. For it to have an appreciable energy density today and to account for all of dark energy, one finds that its mass must be of order of the Hubble constant, namely $m_\phi \sim H_0 \sim 10^{-33}$ eV. The DM axion, on the other end, must be heavy so it  oscillates rapidly enough around its minimum, in order to behave like a dust field and play the role of CDM. If it is to compose all of dark matter today, it has to be oscillating as early as some time before recombination, so that $m_\chi \gtrsim H_\mathrm{rec} \sim 10^{-29}$ eV. However, bounds on axion DM from cosmic growth formation tightens the lower bound to $m_\chi \gtrsim 10^{-20}$ eV \cite{article}. Combining these requirements, we find that our regime of interest is $m_\chi \gg m_\phi \sim H_0$. 

\subsection{Dark matter oscillations}\label{subsec: Dark matter oscillations}

For a DM axion mass at this scale ($m_\chi \gtrsim 10^{-20}$ eV), the fast oscillations at the bottom of its potential make the numerical evaluation of its background cosmic evolution at later times unfeasible. However, as it turns out, the hierarchy of mass scales in the interacting dark sector allows us to frame the heavier axion dynamics in terms of a harmonic oscillator with a (slowly) time-varying mass. To see this, we expand the potential Eq.~\eqref{two-axion potential} around $\tilde{\chi} = \chi - \chi_\mathrm{min}$, where $\chi_\mathrm{min}$ is the field value that minimizes Eq.~\eqref{two-axion potential} along the $\chi$-direction, namely $\frac{\partial V_\mathrm{DS}(\phi, \chi)}{\partial \chi}\Big |_{\chi = \chi_\mathrm{min}} = 0$, which yields the $\phi$ dependent result
\begin{equation}\label{chi_0}
    \frac{\chi_\mathrm{min}}{f_\chi} = \arctan\left[\frac{\beta \sin(\frac{\phi}{f_\phi})}{1+\beta\cos(\frac{\phi}{f_\phi})}\right]\;.
\end{equation}
The parameter $\beta \equiv \frac{\Lambda_{\phi,\chi}^4}{\Lambda_\chi^4}$, given by the ratio between the scales of the mixing and the bare-$\chi$ terms in Eq.~\eqref{two-axion potential}, controls the coupling strength between the two axions. Whenever $\beta = 0$, one recovers the uncoupled lagrangian for two axion-like fields.

Assuming that $\chi$ is heavy enough for it to be close to its minimum at any given time of interest, and assuming $\phi$ is barely moving, we can Taylor expand its potential around $\chi_\mathrm{min}$ to write an effective potential
\begin{equation}
    V_{\mathrm{eff},\chi}(\tilde{\chi}) \sim V_\mathrm{DS}(\phi,\chi_\mathrm{min}) + \frac{1}{2}\frac{\partial^2 V_\mathrm{DS}(\phi, \chi)}{\partial \chi^2}\Big |_{\chi = \chi_\mathrm{min}}{\tilde{\chi}}^2\;,
\end{equation}
where the second derivative of the dark sector potential evaluated at $\chi_\mathrm{min}$ is given by
\begin{equation}\label{potential second derivative at chi_min}
    \frac{\partial^2 V_\mathrm{DS}(\phi, \chi)}{\partial \chi^2}\Big |_{\chi = \chi_\mathrm{min}} = m_\chi^2\mathcal{A}(\phi,\chi_\mathrm{min})\left[1 + \beta\cos\left(\frac{\phi}{f_\phi}\right)\right]\;,
\end{equation}
with 
\begin{equation}\label{mathcalA}
\mathcal{A}(\phi,\chi_\mathrm{min}) = \cos\left(\frac{\chi_\mathrm{min}}{f_\chi}\right) + \frac{\beta\sin\left(\frac{\chi_\mathrm{min}}{f_\chi}\right)\sin\left(\frac{\phi}{f_\phi}\right)}{1 + \beta\cos\left(\frac{\phi}{f_\phi}\right)}\;.
\end{equation}

We will be interested in cases where the coupling is small such that $\beta \lesssim \mathcal{O}(10^{-1})$. In this instance, the $\phi$-dependent minimum of the $\chi$ field given by Eq.~\eqref{chi_0} is at most of order $\arctan(\beta) \sim \beta \sim \mathcal{O}(10^{-1})$, which means that $\mathcal{A}(\phi,\chi_\mathrm{min}) \sim 1$. Therefore, without loss of generality, we approximate the effective potential of $\chi$ around its minimum as
\begin{equation}\label{V eff DM only}
    V_{\mathrm{eff},\chi}(\chi) \approx V_\mathrm{DS}(\phi,\chi_\mathrm{min}) + \frac{1}{2}m_\chi^2(\phi)\chi^2
\end{equation}
with
\begin{equation}\label{V eff DM and m_chi(phi)}
    m_\chi^2(\phi) = m_\chi^2\left[1+\beta\cos(\phi/f_\phi)\right]\;,
\end{equation}
where we have omitted the tilde on $\chi$ from now on.

Hence, for a heavy DM axion at the bottom of its potential, the effect of a small interaction with the DE field is to produce a modulation of its mass, so that its dynamics are effectively described by a quadratic potential subject to a $\phi$-dependent mass term, in a similar fashion to the fermionic DM case coupled to a DE scalar.

In this regime, since $\phi$ is effectively frozen, the first term in Eq.~\eqref{V eff DM only} is a constant, hence the $\chi$ equation of motion is given by
\begin{equation}\label{chi EOM}
    \ddot{\chi} + 3H\dot\chi + m_\chi^2(\phi)\chi = 0\;,
\end{equation}
where dots denote derivatives with respect to cosmic time.

\begin{figure}
    \centering
    \includegraphics[width=\linewidth]{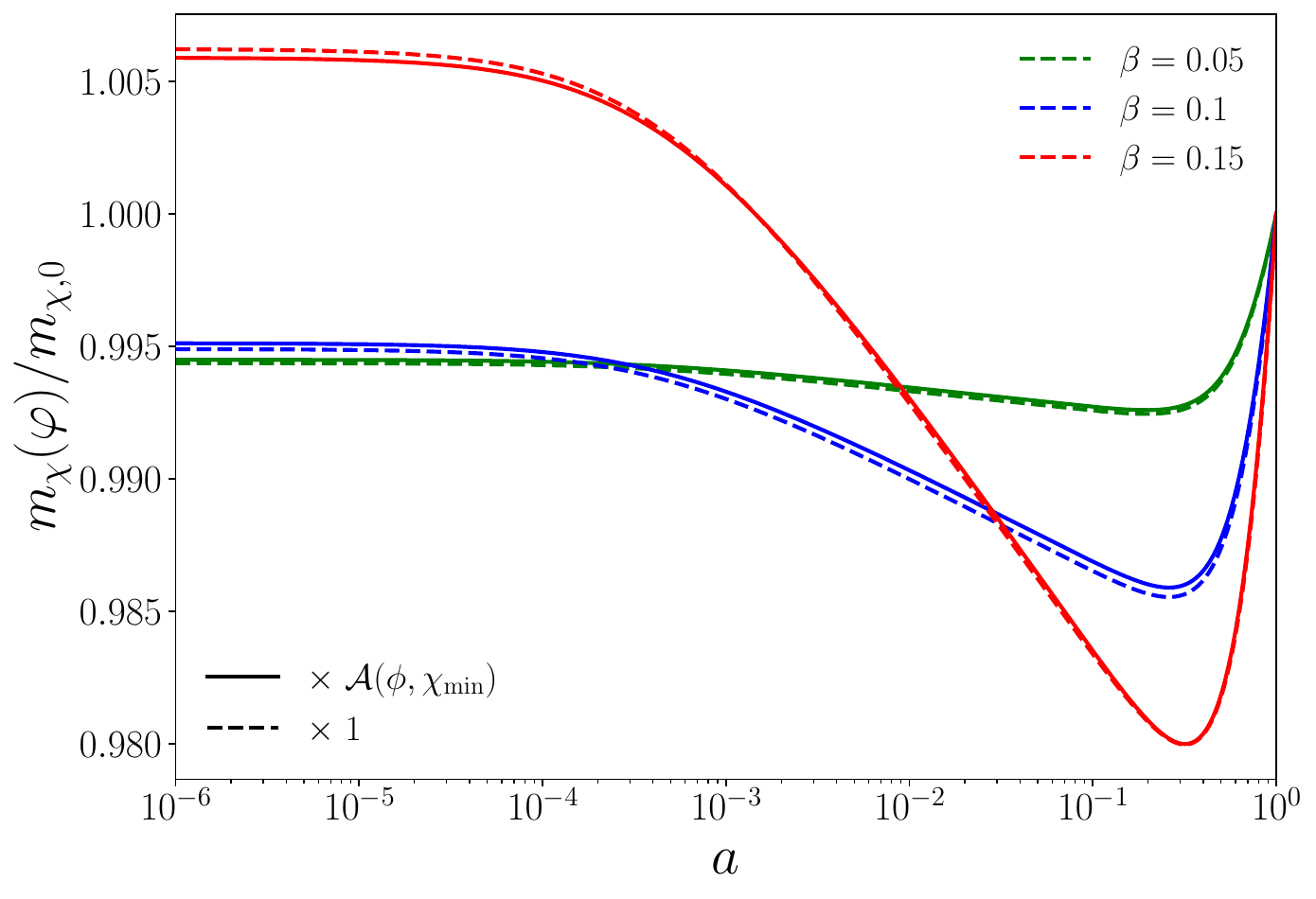}
    \caption{Evolution of the $\phi$-dependent axion DM mass normalized to its present-value, where solid lines use $\mathcal{A}(\phi,\chi_\mathrm{min})$ as given by \eqref{mathcalA} and dashed lines show the approximation used in \eqref{V eff DM and m_chi(phi)}. Here we take $m_\phi = 1.5~H_0$ and $\phi_i = 1.64~f_\phi$, but the validity of the approximation is not sensitive to these parameter choices.}
    \label{fig: m_chi of phi}
\end{figure}

A crucial aspect of such a model is that, since $\phi$ is taken to be very light, in order to behave like dark energy at the present-day, it mustn't evolve much at early times. Thus, we expect $m_\chi(\phi)$ to initially stay approximately constant, before $\phi$ starts rolling down its effective potential. This is shown in Fig. \ref{fig: m_chi of phi} for increasing strength of the coupling $\beta$ and $m_\phi = 1.5 H_0$. We show the evolution of $m_\chi(\phi)$ as given by the exact expression in Eq.~\eqref{potential second derivative at chi_min} (solid curves), as well as the approximation in Eq.~\eqref{V eff DM and m_chi(phi)} (dashed curves), where we take $\mathcal{A}(\phi,\chi_\mathrm{min}) \sim 1$, which provides an accurate description of the $\phi$-dependent DM mass.


Another important feature in Fig. \ref{fig: m_chi of phi} is that for all $\beta > 0$ the DM mass at late-times is smaller than its present-day value, which causes a dip in the associated DM energy density. As it turns out, this is key to producing an apparent phantom-crossing in the effective DE EoS, as we will discuss in Sec. \ref{sec: Background evolution and apparent phantom-crossing}.

In this scenario of a slowly-varying DM mass, the DM axion dynamics are described by oscillating solutions with increasing frequency and an amplitude that scales as $\sim a^{-3/2}$, such that the time averaged DM energy density explicitly behaves like a pressureless fluid, i.e $\rho_\chi \sim V(\chi) \sim \chi^2 \sim a^{-3} $. To see this, it is convenient to factor out the overall scaling of the oscillating $\chi$ field, such that $\chi(t) = a^{-3/2} u(t)$. In terms of $u(t)$, Eq.~\eqref{chi EOM} can be written as
\begin{equation}\label{u EOM}
    \ddot{u} + m_u^2(t) u = 0\;,
\end{equation}
where 
\begin{equation}\label{msqu}
m_u^2(t) = m_\chi^2(\phi(t)) - \frac{9}{4}H^2 - \frac{3}{2}\dot H\;.
\end{equation} 
Eq. \eqref{u EOM} is then the equation of motion of a harmonic oscillator with a time-dependent mass given by Eq. \eqref{msqu}, which, in the most general case, oscillates with a time-varying amplitude and phase. In the case where $m_u(t)$ is slowly varying, we can employ a WKB approximation for the evolution of $u(t)$, (see for example \cite{merzbacher1997quantum})
\begin{equation}\label{u(t) WKB}
    u(t) = \frac{C}{\sqrt{m_u(t)}}\exp{\left[\pm i\int m_u(t) dt\right]}\;,
\end{equation}
for $C$ a constant with dimensions of $\mathrm{[mass]^{3/2}}$. The solution above is valid as long as $\varepsilon \equiv \dot{m}_u/m_u^2$ is much less than unity. 

Since $m_\chi \gg H$, $m_u(t) \approx m_\chi(\phi(t))$, meaning that the $\varepsilon \ll 1$ condition is tied to the slow-roll evolution of $\phi$, so that the DM axion mass is approximately constant. In this regime, the WKB solution for $u(t)$, Eq.~\eqref{u(t) WKB} signals that the DM axion {\color{blue} $\chi$} oscillates with increasing frequency and an amplitude $A(t)$ that scales as $A(t) = \ C a^{-3/2}/\sqrt{m_u(t)}$. 
Therefore, we assign the energy density of the $\chi$ field as the time-average energy density of its WKB solution, which effectively captures the evolution of the field's envelope, namely
\begin{equation}\label{rhochi-wkb}
    \rho_\chi = \langle \rho_{\chi,\mathrm{WKB}} \rangle = \frac{\langle \dot{\chi}^2\rangle}{2} + \frac{m_\chi^2(\phi)}{2}\langle\chi^2\rangle\;.
\end{equation}

As long as $\varepsilon \ll 1$, the oscillating solution for $u(t)$ -- and, consequently, $\chi(t)$ -- stays on the WKB branch such that $\rho_\chi$ follows the dust-like evolution.

In Appendix \ref{app:A}, we perform a comparison between the full numerical computation of the two-axion dynamics and the WKB approximation, where we show that the approximate solution holds in the $\chi$ oscillatory regime, with $\varepsilon$ decreasing as $m_\chi$ increase. As we mentioned earlier, this is unfeasible numerically when using the required hierarchy between the DM and DE masses, so in Appendix \ref{app:A} we show the full evolution for the case where $m_\chi = 6.7\cdot 10^{2} ~m_\phi$. Our aim is to demonstrate how well the approximate solution holds in the $\chi$ oscillatory regime.

\subsection{Coupled quintessence framework}

An important consequence of the WKB evolution of $\chi$ is that its energy density becomes proportional to its mass, as in the standard coupled quintessence framework. Considering the time averaged energy density $\rho_\chi$, Eq.~\eqref{rhochi-wkb}, we can write $\langle\chi^2\rangle = A^2(t)/2$, hence  Eq.~\eqref{rhochi-wkb} becomes
\begin{equation}
    \rho_\chi \sim \frac{a^{-3}}{4}C^2 m_\chi(\phi)\left[1+\varepsilon^2 + 3\frac{H}{m_\chi(\phi)}\varepsilon + \frac{9}{4}\frac{H^2}{m_\chi^2(\phi)} \right]\;,
\end{equation}
where recall we have taken $m_u(t) \approx m_\chi(\phi)$ since $m_\chi(\phi) \gg H$ during the oscillatory regime. Therefore, all the variable terms inside the above bracket are suppressed either by $\varepsilon$ or $H/m_\chi(\phi)$, such that $\rho_\chi \sim \frac{a^{-3}}{4}C^2 m_\chi(\phi)$ to leading order. Thus, the DM energy density normalized to its present-day value is
\begin{equation}\label{rho_chi scaling}
    \frac{\rho_\chi}{\rho_{\chi,0}} = \frac{m_\chi(\phi)}{m_{\chi,0}}a^{-3}\;,
\end{equation}
which follows the standard CDM evolution law with, from Eq.~\eqref{V eff DM and m_chi(phi)}, a $\phi$ modulation proportional to $\sqrt{1 + \beta\cos(\phi/f_\phi)}$ \footnote{Since we are working in the small coupling regime, the square-root is always going to be real in the $0<\phi/f_\phi<\pi$ region.}. Note that the form of Eq.~\eqref{rho_chi scaling} implies that the bare DM mass, $m_\chi$, drops out of the dynamics in the fluid approach. It is a common feature of coupled quintessence models. 

Eq. \eqref{rho_chi scaling} is precisely the DM scaling one gets when working in the coupled quintessence framework, where the DE-DM coupling is described by two interacting fluids \cite{Pourtsidou:2013nha}. As it turns out, it is the solution of a sourced CDM continuity equation \cite{Brookfield:2007au,Xia_2009}
\begin{equation}\label{CDM continuity eq}
    \rho'_\chi + 3 \mathcal{H} \rho_\chi = Q(\phi)\;,
\end{equation}
where
\begin{equation}\label{Q_of_phi}
     Q(\phi) = \gamma(\phi) \rho_\chi(\phi) \phi'\;,\hspace{0.3cm}\gamma(\phi)\equiv \frac{d \ln m_\chi(\phi)}{d \phi}\;,
\end{equation}
$\mathcal H = a H$ and primes denote derivatives with respect to conformal time $d\tau = dt/a$. The quantity $Q(\phi)$ is the coupling function between the two fluids and plays an important role in the dynamics of coupled scalar field models. If $Q>0$, energy flows from the $\phi$ scalar field (DE) to the $\chi$ (DM) scalar field component. In Section \ref{sec: Background evolution and apparent phantom-crossing}, we will analyze how the late-time behavior of $Q(\phi)$ is key, allowing us to produce the energy balance necessary to fit current observational data.

In terms of the dynamical DM mass given in Eq.~\eqref{V eff DM and m_chi(phi)}, the $Q(\phi)$ and $\gamma(\phi)$ functions are explicitly given by
\begin{equation}\label{Q and gamma of phi explict}
    \begin{split}
        Q(\phi) & = -\frac{\beta\sin(\phi/f_\phi)}{2 f_\phi\sqrt{1+\beta\cos(\phi/f_\phi)}}\rho_{\chi,0}a^{-3}\phi'\hspace{0.3cm}\mathrm{and}\\
        \gamma(\phi)& = -\frac{\beta\sin(\phi/f_\phi)}{2 f_\phi^2[1+\beta\cos(\phi/f_\phi)]}\;.
    \end{split}
\end{equation}

Per energy conservation, the DE energy density of $\phi$ will obey a continuity equation with a source term of the same magnitude and opposite sign, which is associated with the Klein-Gordon equation for the DE scalar field 
\begin{equation}\label{KG eq}
    \phi'' + 2 \mathcal{H}\phi' + a^2\frac{d V_\phi (\phi)}{d \phi} = -a^2\frac{Q(\phi)}{\phi'}\; .
\end{equation}
Eqs. \eqref{CDM continuity eq} and \eqref{KG eq} are supplemented by the Friedmann equation for the conformal Hubble rate
\begin{equation}\label{Friedmann eq}
    \mathcal{H}^2 \equiv \left(\frac{a'}{a}\right) = \frac{8\pi G}{3}\left( \rho_r + \rho_b + \rho_\chi + \rho_\phi \right)\;,
\end{equation}
where the baryonic and radiation energy densities follow their standard scalings as $\rho_b = \rho_{b,0}a^{-3}$ and $\rho_r = \rho_{r,0}a^{-4}$, respectively.

Also, due to the energy-momentum exchange between the DE and DM fluids, the evolution of linear perturbations for the coupled system will also be modified in comparison to the uncoupled case. Therefore, assuming a coupling function of the form Eq.~\eqref{Q_of_phi}, the perturbation equations for the DM density contrast $\delta_\chi \equiv \delta\rho_\chi/\rho_\chi$ and the velocity divergence $\theta_\chi \equiv \partial_i v^i_\chi$ are written in the synchronous gauge as \cite{Brookfield:2007au,Khoury:2025txd} 
\begin{equation}\label{perturbations CDM}
    \begin{split}
        \delta_\chi' = -\theta_\chi - \frac{h'}{2} + \gamma(\phi)\delta\phi' + \gamma_{,\phi}\phi'\delta\phi\;, \\
        \theta_\chi' = - \mathcal{H}\theta_\chi - \gamma(\phi)\phi'\theta_\chi + \gamma(\phi)k^2\delta\phi\;,
    \end{split}
\end{equation}
while the perturbed KG equation for $\delta\phi$ reads
\begin{equation}\label{perturbations KG}
    \delta\phi'' = - 2\mathcal{H}\delta\phi' - k^2\delta\phi - \frac{h'}{2}\phi' - V_{,\phi\phi}\delta\phi - a^2[\rho_\chi\gamma_{,\phi}\delta\phi + \gamma\rho_\chi\delta_\chi]\;, 
\end{equation}
with $h$ being the metric trace perturbation in the synchronous gauge.

\section{Background evolution and apparent phantom-crossing}\label{sec: Background evolution and apparent phantom-crossing}

As discussed in Sec.~\ref{sec:1}, the effect of a phantom DE component in the cosmic expansion rate can be mimicked by having a smaller DM energy density in the recent past compared to the standard CDM expectation. Assuming that the present-day densities are unchanged, this can only be the case if $\rho_\chi$ redshifts slower than the standard $\sim a^{-3}$ law, i.e, there is an extra injection of energy in the DM component that dillutes the effect of cosmic expansion. In the context of coupled quintessence, this is the case whenever $Q(\phi) > 0$ in the recent past, which becomes a requirement in order to obtain a late-time apparent phantom-crossing. 

As we shall now demonstrate, this behavior of the coupling function is completely tied to the dynamics of the DE scalar, which both depends on the shape of its effective potential, and takes into account the background coupling to the DM axion. In this sense, using Eq.~\eqref{rho_chi scaling} and Eq.~\eqref{Q_of_phi}, Eq. \eqref{KG eq} can be written as 
\begin{equation} \label{phieqofmotion}
    \phi'' + 2\mathcal{H}\phi' + a^2\frac{d V_{\mathrm{eff},\phi}(\phi)}{d \phi} = 0\; ,
\end{equation}
with $V_{\mathrm{eff},\phi}(\phi)$ given by 
\begin{equation}\label{V_eff(phi)}
    \begin{split}
        V_{\mathrm{eff},\phi}(\phi)  &= V_\phi(\phi) + \rho_\chi(\phi)\\
        & = \Lambda^4_\phi\left[1 - \cos\left( \frac{\phi}{f_\phi} \right) \right]\\& + \rho_{\chi,0}a^{-3}\sqrt{1+\beta\cos\left(\frac{\phi}{f_\phi}\right)}\;.
    \end{split}
\end{equation}

At sufficiently early times, the CDM density $\rho_\chi(\phi)$ is large compared to the energy stored in the DE scalar field potential $V(\phi)$, so the $\rho_\chi$ term dominates the effective potential. Also, due to Hubble friction, $\phi$ initially stays approximately frozen, so we can drop the second-derivative term in Eq.~\eqref{phieqofmotion}, leading to an early time approximation for $\phi'$ 
\begin{equation}\label{phi_prime approx}
     \phi'  \sim - \frac{a^2}{2\mathcal{H}}\frac{d \rho_\chi(\phi)}{d \phi} = - \frac{a^2}{2\mathcal{H}}\rho_\chi(\phi) \gamma(\phi) \; ,
\end{equation}
where, in the last equality, we have substituted Eq.~\eqref{rho_chi scaling}. Inserting $\phi'$ from the above expression back into $Q(\phi)$ as in Eq.~\eqref{Q_of_phi}, we find
\begin{equation}\label{Q of phi approx}
    Q(\phi) \sim -\frac{a^2}{2\mathcal{H}}\rho_\chi^2(\phi) \gamma^2(\phi)\; .
\end{equation}
The right-hand side of the above expression is, by construction, negative. Therefore, we conclude that the initial coupling function always satisfies $Q < 0$, signaling an early energy exchange from DM to DE. Thus, in order to satisfy the late-time requirement of $Q > 0$, the scalar field dynamics of $\phi$ and $\chi$ have to evolve in such a way that $Q \propto \gamma(\phi) \phi'$ changes sign. As it turns out, in the $0<\phi/f_\phi < \pi$ region and for $\beta \sim\mathcal{O}(10^{-1})$, $\gamma(\phi)$ as given in Eq.~\eqref{Q and gamma of phi explict} is always negative, so the sign change of $Q(\phi)$ depends on the change of $\phi'$ from initial positive to negative values.

The initial behaviour of the field's derivative and its coupling function also determines the scaling of its energy density, which is given by the sum of the kinetic and potential terms, namely 

\begin{figure}
    \centering
    \includegraphics[width=\linewidth]{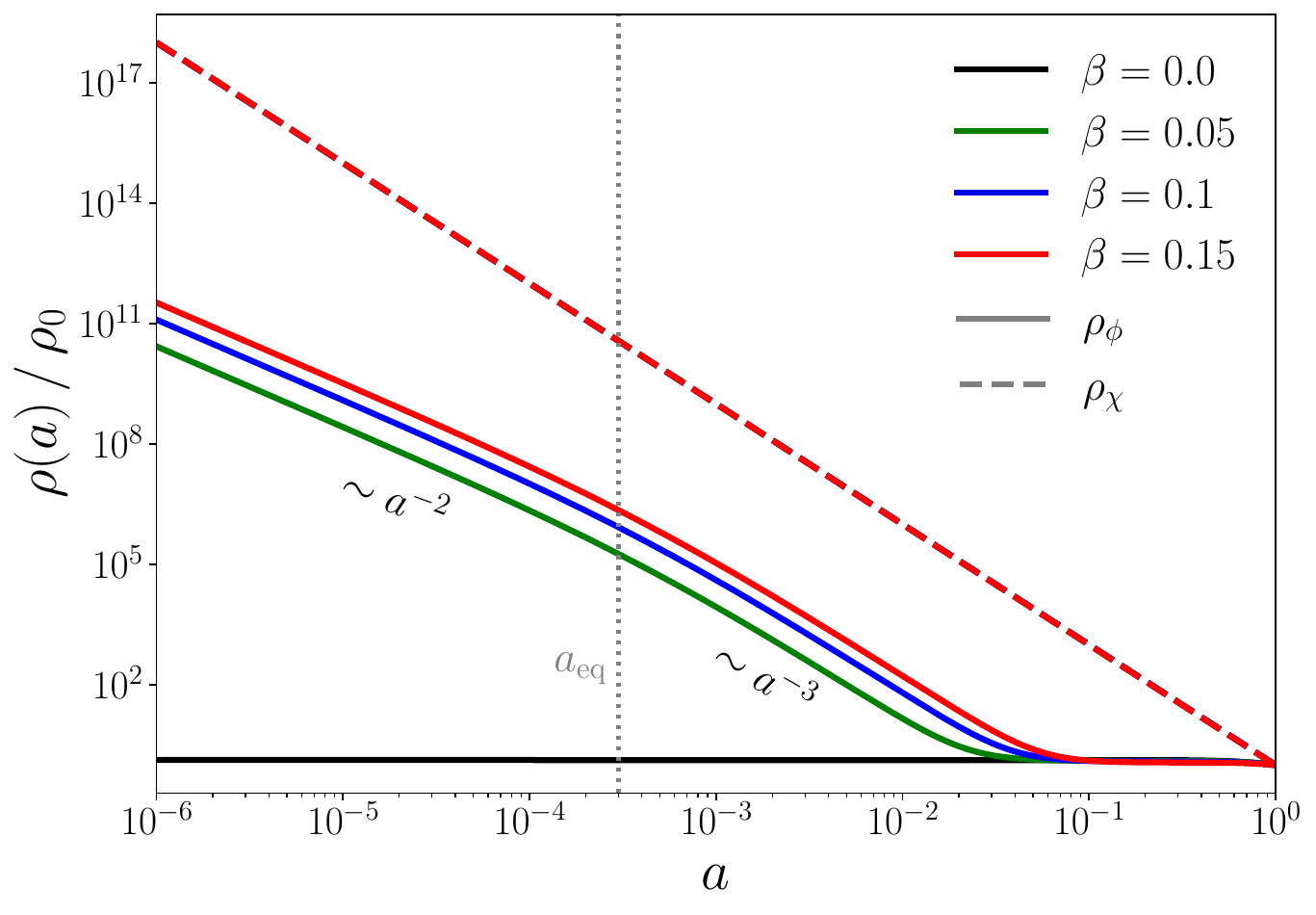}
    \caption{Evolution of the normalized DE (solid, Eq.~\eqref{def:rhophi}) and DM (dashed, Eq.~\eqref{rho_chi scaling}) energy densities for different coupling strengths. Here we take $m_\phi = 1.5~H_0$ and $\phi_i = 1.64~f_\phi$. The vertical gray dotted line represents the scale factor of radiation-matter equality. 
    \label{fig: rho_DE and rho_DM}}
\end{figure}

\begin{equation}\label{rho_scaling}
    \rho_\phi = \frac{\phi'^2}{2 a^2} + V_\phi 
    \sim \frac{\gamma^2(\phi)}{8}\frac{\rho_\chi^2(\phi)}{H^2} + V_\phi\;,
\end{equation}
where we substituted $H = \mathcal{H}/a$.

At early times, the field stays approximately frozen, so that $\gamma(\phi)$ and $V_\phi$ are slowly-varying functions. The scale-dependence of $\rho_\phi$ is then encoded in the first term in Eq.~\eqref{rho_scaling}, 
hence, by approximating $H^2 \sim \rho_r/M_P^2 = \Omega_r H_0^2 a^{-4}$ and $H^2 \sim \rho_{\chi}/M_P^2 = \Omega_\chi H_0^2 a^{-3}m_\chi(\phi)/m_{\chi,0}$ during the radiation-dominated (RD) and matter-dominated (MD) epochs, respectively, we find 
\begin{equation}\label{rho_scaling_RD_and_MD}
    \rho_\phi(a) \sim \begin{cases}
     \frac{\gamma^2(\phi)}{8}\frac{m_\chi^2(\phi)}{m^2_{\chi,0}}\frac{\Omega_{\chi,0}^2}{\Omega_{r,0}}H_0^2 M_P^4 a^{-2}\text{(RD)} \\
     \frac{\gamma^2(\phi)}{8}\frac{m_\chi(\phi)}{m_{\chi,0}}\Omega_{\chi,0}H_0^2 M_P^4 a^{-3}\hspace{0.5cm}\text{(MD)}\;.
    \end{cases}
\end{equation}

Therefore, the coupled quintessence field tracks the DM component during the MD epoch -- with $\rho_\phi/\rho_\chi \sim \gamma^2(\phi)M_P^2/8$ -- and scales as $a^{-2}$ during RD, which is displayed as the colored solid lines in Fig. \ref{fig: rho_DE and rho_DM}. As shown in \cite{Archidiacono:2022iuu}, this behavior is quite general and independent of the particular choice of the coupling function, since the only assumption we have made is that the energy density of $\phi$ is initially much smaller than the DM energy density and that $\phi$ is initially frozen due to Hubble friction. On the other hand, the DM density $\rho_\chi$, defined in Eq.~\eqref{rho_chi scaling} receives a small modulation proportional to $m_\chi(\phi)$ but still decays as dust irrespectively of the coupling strength, represented by the superimposed dashed lines. 

We now turn our attention to the overall motion of the DE axion $\phi$, which follows directly from tracking the minimum of its effective potential, Eq.~\eqref{V_eff(phi)}. The critical points of the effective potential are found by evaluating $\frac{d V_{\mathrm{eff},\phi}(\phi)}{d\phi}\Bigg|_{\phi=\phi_\mathrm{crit}} = 0$, which we find is equivalent to solving
\begin{equation}\label{V_eff critical points}
     \left(\frac{\Lambda_\phi^4}{f_\phi} - \frac{\rho_{\chi,0}a^{-3}\beta}{2 f_\phi\sqrt{1+\beta\cos\varphi_\mathrm{crit}}}\right)\sin\varphi_\mathrm{crit} = 0 \;,
\end{equation}
where we define $\varphi\equiv\phi/f_\phi$. In general, the term in parentheses will not vanish, so the critical points are located at $\sin\varphi_\mathrm{crit} = 0$, or $\varphi_\mathrm{crit} = n\pi$ for $n \in \mathbb{Z}$. Whether these points correspond to a minimum or a maximum depends on the overall sign of the effective potential second derivative which is given by
\begin{equation}
    \begin{split}
        \frac{d^2 V_{\mathrm{eff},\phi}(\phi)}{d\phi^2}= & m_\phi^2\cos\varphi  -\frac{\rho_{\chi,0}a^{-3}\beta^2\sin^2\varphi}{4 f_\phi^2(1+\beta\cos\varphi)^{3/2}}\\
        &-  \frac{\rho_{\chi,0}a^{-3}\beta\cos\varphi}{2 f_\phi^2\sqrt{1+\beta\cos\varphi}}\;,
    \end{split}
\end{equation}
where $m_\phi = \Lambda_\phi^4/f_\phi^2$. Evaluating the above expression for the cases of $n=0$ and $n=1$, we find
\begin{equation}
    \begin{split}
        \frac{d^2 V_{\mathrm{eff},\phi}(\phi)}{d\phi^2}\Bigg|_{\phi=0} & = m_\phi^2 - \frac{\rho_{\chi,0}a^{-3}\beta}{2 f_\phi^2\sqrt{1+\beta}}\hspace{0.3cm}\mathrm{and}\\
        \frac{d^2 V_{\mathrm{eff},\phi}(\phi)}{d\phi^2}\Bigg|_{\phi=\pi} & = -m_\phi^2 + \frac{\rho_{\chi,0}a^{-3}\beta}{2 f_\phi^2\sqrt{1-\beta}}\;.
    \end{split}
\end{equation}

Substituting $\rho_{\chi,0} = \Omega_c H_0^2 M_P^2$, the point $\varphi = 0$ will be a minimum when
\begin{equation}
    a > a_1 = \left( \frac{\Omega_c H_0^2 M_P^2}{2 m_\phi^2 f_\phi^2}\frac{\beta}{\sqrt{1+\beta}} \right)^{1/3}
\end{equation}
while $\varphi = \pi$ is a minimum when
\begin{equation}
    a < a_2 = \left( \frac{\Omega_c H_0^2 M_P^2}{2 m_\phi^2 f_\phi^2}\frac{\beta}{\sqrt{1-\beta}} \right)^{1/3}\;.
\end{equation}

For $\Omega_c\sim 0.26$, $\beta\sim0.1$, $m_\phi \sim H_0$ and $f_\phi \lesssim M_p$, we find that the redshift corresponding to $a_1$ and $a_2$ is of order $z\sim 1$, with a difference of $\Delta z \lesssim 0.1$ between them. During this small time-span when $a_1<a<a_2$, both $\varphi=0$ and $\varphi=\pi$ are minima and the effective potential develops a transition hilltop located at
\begin{equation} \label{def:phi-crit}
    \varphi_\mathrm{crit} = \arccos\left\{\frac{1}{\beta}\left[ \left(\frac{\Omega_c H_0^2 M_P^2\beta}{2 m_\phi^2 f_\phi^2 a^3}\right)^2 -1 \right]\right\}\;,
\end{equation}
which is associated with the vanishing parentheses in Eq.~\eqref{V_eff critical points}.

\begin{figure*}
    \centering
    \includegraphics[width=\linewidth]{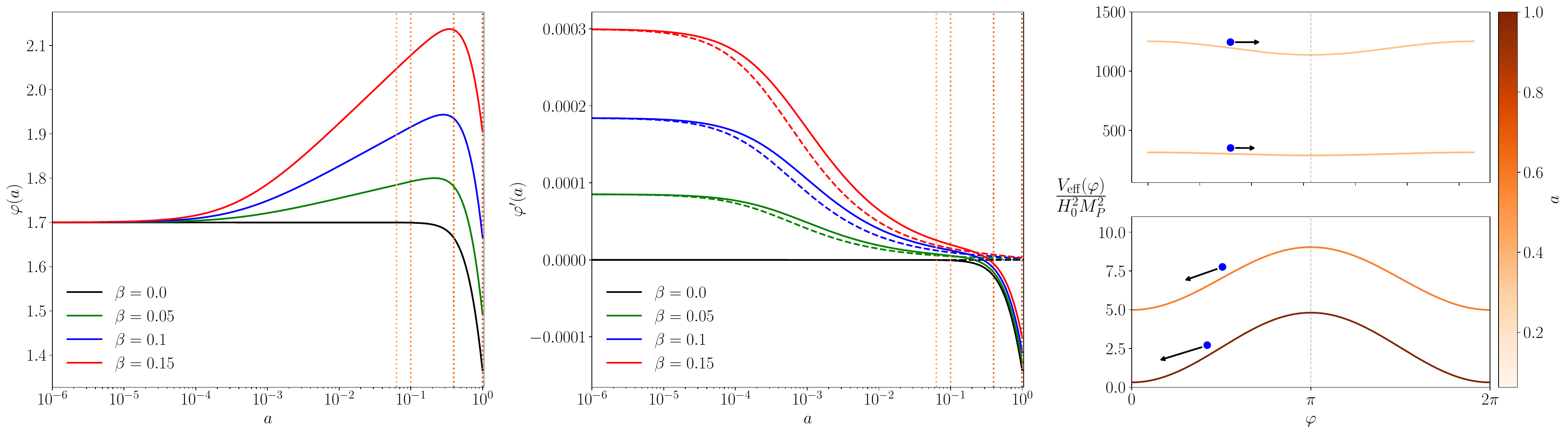}
    \caption{Background dynamics of the DE axion $\varphi$ (left) and its conformal derivative $\varphi'$ (middle) for the uncoupled quintessence case (black) as well as three representative cases of a non-zero coupling constant $\beta$. In the right most panel, we show the evolution of the effective potential for the $\beta = 0.1$ case across four distinct values of the scale factor, which are depicted as dashed vertical lines in the first two panels. We also show the approximate early time evolution of $\varphi '$, according to Eq. \eqref{phi_prime approx}, as colored dashed lines in the middle panel. Here we also take $m_\phi = 1.5~H_0$ and $\phi_i = 1.64~f_\phi$.}
    \label{fig: varphi, varphi_prime and V_eff}
\end{figure*}

In Fig. \ref{fig: varphi, varphi_prime and V_eff} we show the overall DE dynamics {\color{blue} of $\phi$} for $m_\phi/H_0 = 1.5$. The right-most panel shows the evolution of the effective potential Eq.~\eqref{V_eff(phi)} with $\beta = 0.1$ for four distinct values of the scale-factor -- which increases upward in the color bar -- highlighting the shift of the minimum from $\varphi_\mathrm{crit} = \pi$ at early times to $\varphi_\mathrm{crit} = 0$ in the late universe. The arrows indicate the magnitude and direction of the field's conformal derivative, $\varphi'$ at these four moments in time. In particular we note that it changes direction as required from Eq.~\eqref{Q of phi approx}.

We see this directly in the left most panel which displays the background evolution of the normalized DE axion $\varphi$ for the uncoupled case ($\beta = 0$) in black, as well as for three distinct non-zero values of $\beta$. In all cases, the field is initially frozen due to Hubble damping, but the evolution from matter domination onward changes depending on the coupling strength. For these coupled cases, the field initially begins its motion from the damped state by tracking the early minimum at $\varphi_\mathrm{crit} = \pi$. When $a > a_2$, its trajectory is reversed, on the way to the new overall minimum at  $\varphi_\mathrm{crit} = 0$. 

This behaviour is supported by the middle panel which explicitly shows the evolution of $\varphi'$. For the non-zero $\beta$, the derivative is initially locked in the positive attractor of Eq. \eqref{phi_prime approx}, represented by the dashed curves. As the approximation starts to fail, $\varphi '$ switches sign after the effective potential transition, reflecting the position change of the minimum. In both of the first two panels, as colored vertical dotted lines, we show the scale factors associated with each of the four values of $V_{\mathrm{eff},\phi}(\varphi)$ displayed in the right panel.

\begin{figure*}
    \centering
    \includegraphics[width=\linewidth]{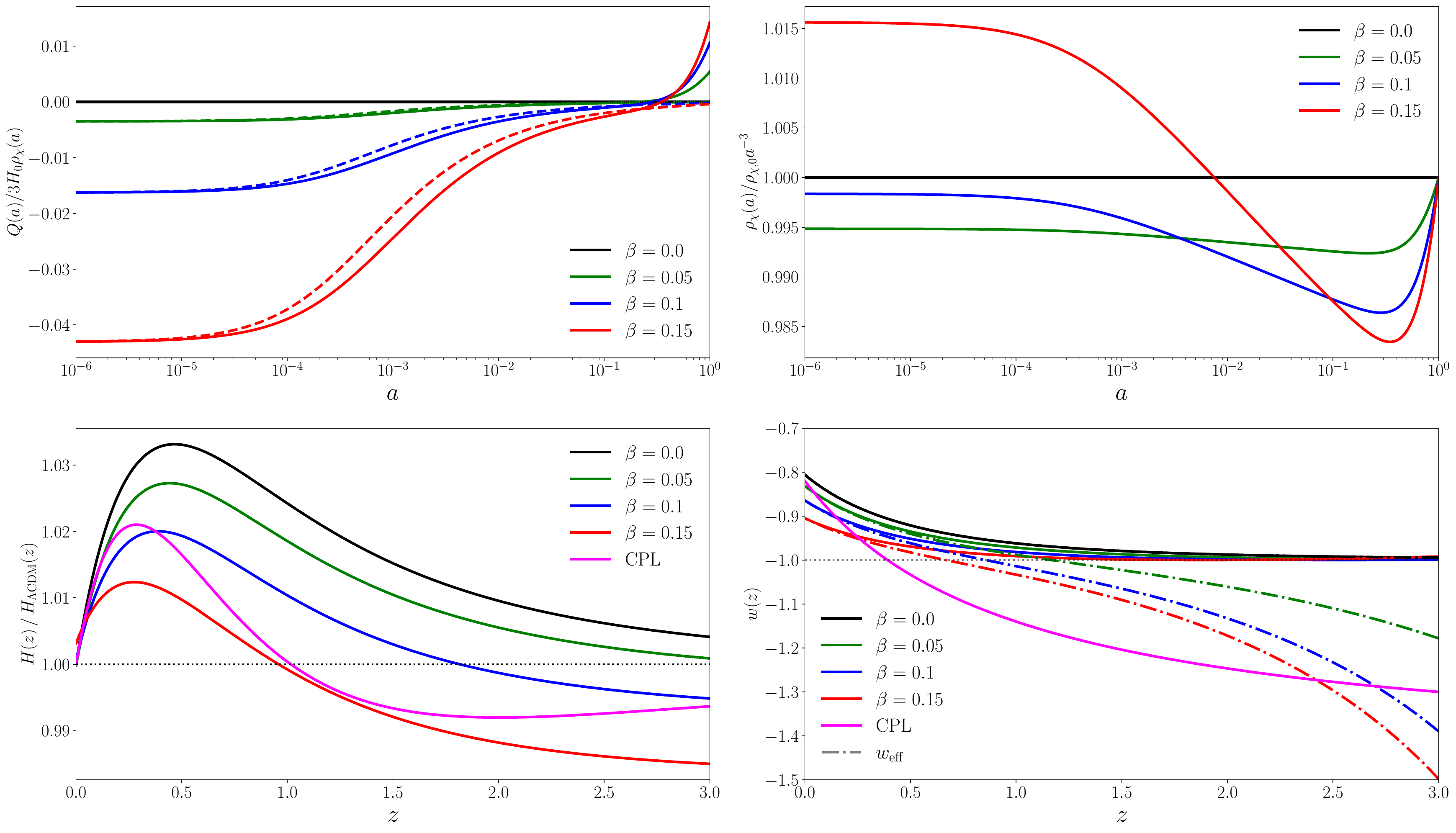}
    \caption{Background evolution of the dimensionless coupling function Eq.~\eqref{Q and gamma of phi explict} (top left), the DM energy density \eqref{rho_chi scaling} normalized to standard CDM (top right), the Hubble rate normalized to its $\Lambda$CDM from (bottom left) and the DE equation of state (bottom right). In all panels, black curves denote the uncoupled case while green, blue and red curves represent $\beta = 0.05,0.1, 0.15$, respectively. The dashed curves depict the approximate early-time coupling function given by Eq. \eqref{Q of phi approx} in the upper left panel and the apparent phantom-crossing DE EoS Eq.~\eqref{w_eff} in the bottom right, respectively. The pink curves in the bottom two panels show the CPL functions assuming the CMB+DESI DR2+Pantheon+ best-fit from \cite{DESI:2025zgx}.}
    \label{fig: background quantities}
\end{figure*}

As discussed earlier around Eq.~\eqref{Q of phi approx}, this general behavior of the DE scalar $\varphi$ in coupled quintessence scenarios -- namely a sign change in its conformal derivative -- is key for producing the late-time observables consistent with the findings of the DESI Collaboration. This is highlighted in the panels in Fig. \ref{fig: background quantities}. In the upper left panel, we show the evolution of the dimensionless coupling function $Q(a)$ for the same representative cases of $\beta$ as in Fig. \ref{fig: varphi, varphi_prime and V_eff}. Solid curves denote $Q(a)$ in its full generality -- Eq. \eqref{Q and gamma of phi explict}, while dashed curves are the attractor approximation given by Eq. \eqref{Q of phi approx}, which are shown to provide an accurate description at early times. As a consequence of the sign-change of $\varphi '$, the coupling function also changes from negative to positive values in the late universe, which results in less energy density in the DM component compared to the usual CDM scaling, shown in the upper right panel. As a result, this percent-level deviation in $\rho_\chi$ in the coupled case brings the late-time expansion rate closer to the preferred behavior of the CPL parameterization, as seen in the bottom left panel.

Finally, all of these effects can be summarized as an apparent dark energy equation of state, which takes into account what would be the expansion rate had one assumed the standard CDM energy density scaling of $\rho_\chi \sim a^{-3}$. To this end, we take that the dark sector contribution to the Hubble rate, given by the sum $\rho_\phi + \rho_\chi $, can be equally expressed by the sum of an effective dark energy contribution and a standard CDM component, namely $\rho_\mathrm{eff}+\rho_\mathrm{CDM}$, with $\rho_\mathrm{CDM} = \rho_{\chi,0}a^{-3}$. Thus, equating the two, $\rho_\phi + \rho_\chi = \rho_\mathrm{eff}+\rho_\mathrm{CDM}$ and solving for $w_\mathrm{eff} = p_\mathrm{eff}/\rho_\mathrm{eff}$, we find
\begin{equation}\label{w_eff}
    w_\mathrm{eff}(a) = \frac{w_\phi(a)}{\left[\frac{m_\chi(\phi)}{m_{\chi,0}}-1\right]\frac{\rho_{\chi,0}a^{-3}}{\rho_\phi}+1}\;,
\end{equation}
where $w_\phi = p_\phi/\rho_\phi$ is the usual DE scalar field EoS, with $p_\phi = \dot{\phi}^2/2 - V_\phi(\phi)$ the DE pressure and we have substituted $\rho_\chi$ according to Eq.~\eqref{rho_chi scaling}. Both $w_\phi$ (solid) and $w_\mathrm{eff}$ (dashed) are plotted in the bottom right panel of Fig. \ref{fig: background quantities}. While the fiducial DE EoS stays bounded above $-1$, the effective EoS crosses to the $w_\mathrm{eff} < -1$ region at low redshifts, in a similar fashion to the CPL best-fit curve given in pink. Inspection of Eq. \eqref{w_eff} shows that this is achieved for $w_\phi(a)\sim -1$ whenever $m_\chi(\phi) < m_{\chi,0}$, in agreement with Fig. \ref{fig: m_chi of phi} for $\beta > 0$.

Therefore, it is clear that the coupled quintessence scenario can explain the current preferred behavior of the expansion rate without the need to invoke a dark energy component that crosses to the phantom region at low redshifts. With that in mind, we now turn our attention to a statistical analysis with current cosmological data in order to constrain the parameter space of the two-axion scenario and infer its observational validity.

\section{Data Analysis}
\subsection{Methodology and Observational Data}\label{sec:3}

We compute the theoretical predictions of the two-axion model using a modified version of the Cosmic Linear Anisotropy Solving System (\texttt{CLASS}) Boltzmann solver \cite{lesgourgues2011cosmiclinearanisotropysolving,Diego_Blas_2011}, where the interacting dynamics of the coupled quintessence framework have been implemented in both the background and perturbations modules, according to Eqs. \eqref{CDM continuity eq}, \eqref{KG eq} and \eqref{perturbations CDM}, \eqref{perturbations KG}, respectively. We work in the spatially flat limit of the FLRW metric and we fix the effective number of relativistic degrees of freedom to $N_\mathrm{eff} = 3.046$, as well as the sum of neutrino masses to $\sum m_\nu = 0.06$ eV.

For the statistical inference, we use the publicly available \texttt{Cobaya} sampler \cite{Torrado:2020dgo} with the \texttt{Polychord} nested sampler extension \cite{Handley:2015fda, Handley_2015} for accurate computation of the models' Bayesian evidences. We analyze the samples with \texttt{GetDist} \cite{Lewis:2019xzd} and obtain the combined likelihood $\chi^2$ with the \texttt{Py-BOBYQA} minimizer package \cite{10.1145/3338517}. 

We parametrize the initial DE field value in terms of its displacement from the minimum of the effective potential, namely $\delta_i \equiv \varphi_i - \varphi_\mathrm{crit}$, where $\varphi_\mathrm{crit}$ is defined in Eq.~\eqref{V_eff critical points}.  In the coupled axion case, the initial field value deep in the radiation dominated epoch is located at $\varphi_\mathrm{crit} = \pi$, whereas $\varphi_\mathrm{crit} = 0$ when $\beta = 0$, recovering the standard uncoupled case. We leave $\delta_i$ as a free parameter, varying it from $\delta_i \in [0,\pi/2]$, to ensure that the field starts close to its potential minimum, avoiding the issues of fine-tuning commonly present in standard axion dark energy.

For the base $\Lambda$CDM parameters, we sample on $H_0, \omega_b, \log A_s, n_s, \tau_\mathrm{reio}$ and $\Omega_c$. The physical present-day CDM density parameter in the coupled quintessence case becomes a derived parameter and is given by evaluating Eq.~\eqref{rho_chi scaling} today, so that $\Omega_{c,\mathrm{phys}} = \Omega_c m_\chi(\varphi_0)/m_\chi$, with $m_\chi(\varphi_0)/m_\chi = \sqrt{1+\beta\cos\varphi_0}$  and $\varphi_0$ -- the present-day field value -- obtained after numerically solving the background equations of motion. In this setting, the DE axion decay constant $f_\phi$ is also a derived parameter and is fixed by requiring that $\sum \Omega_{i,\mathrm{phys}} = 1$ inside the \texttt{CLASS} shooting routine. Moreover, we limit ourselves to $\log(m_\phi/H_0) \leq 0.35$ since larger values for the DE axion mass make the field oscillate around the initial minimum before the effective potential transition. The full prior list for all the sampled parameters are given in Table \ref{tab:priors}.

\begin{table}[ht]
  \centering
  \begin{tabular}{lc}
  \hline\hline
  Parameter & Prior range \\
  \hline
  $\log(10^{10} A_\mathrm{s})$ & $[2.5,\, 3.5]$       \\
  $n_\mathrm{s}$               & $[0.8,\, 1.2]$       \\
  $H_0$                        & $[50,\, 80]$         \\
  $\omega_b$      & $[0.018,\, 0.026]$   \\
  $\tau_\mathrm{reio}$         & $[0.01,\, 0.12]$     \\
  $\Omega_c$      & $[0.2,\, 0.4]$       \\
  \hline
  $\beta$                      & $[0,\, 0.15]$        \\
  $\log(m_\phi/H_0)$           & $[-1.5,\, 0.35]$     \\
  $\delta_i$                   & $[0.01,\, 1.57]$     \\
  \hline\hline
  \end{tabular}
  \caption{Uniform prior ranges adopted for the base $\Lambda$CDM parameters and the parameters relevant to the interacting axion model.}
  \label{tab:priors}
\end{table}
For the statistical analysis, we employ data from CMB, BAO and SNIa observations according to the following surveys:

\begin{itemize}
    \item \textbf{CMB}: For the power spectra of temperature and polarization anisotropies, we use a combination of both the Legacy 2018 release of the Planck satellite \cite{Planck:2019nip} as well as the Data Release 6 of the ground-based ACT telescope \cite{Qu:2023spj}.

    For low multipoles ($\ell < 30$), we use the \texttt{Commander} and \texttt{SimAll} likelihoods for the TT and EE spectra, respectively. At high multipoles ($\ell \geq 30$), we employ the \texttt{CamSpec} likelihood derived from the Planck PR4 \texttt{NPIPE} maps, which replaced the previous 2018 data release with a new processing pipeline \cite{Rosenberg:2022sdy}.

    Moreover, we use gravitational lensing CMB data from a combination of the Planck PR4 lensing reconstruction \cite{Carron:2022eum} and ACT DR6 \cite{Madhavacheril:2023oyr}, providing more accurate constraints from growth of structure observations.

    \item \textbf{BAO}: We use BAO data from the DESI second Data Release (DR2) \cite{DESI:2025zgx}, which covers the redshift range of  $0.295 \leq z \leq 2.33$. The data is obtained from the two-point correlation function of distinct groups of tracers, namely the Bright Galaxy Sample (BGS), Luminous Red Galaxies (LRGs), Emission Line Galaxies (ELGs), Quasars (QSOs), and the Lyman-$\alpha$ (Ly$\alpha$) forest.

    The final data is released in terms of the Hubble distance $D_H/r_d$, the transverse distance $D_M/r_d$ and the volume-averaged distance $D_V \equiv (z D_M^2 D_H)^{1/3}$, all normalized to the sound-horizon at the drag epoch.

    \item \textbf{SNIa}: In the following, we perform three separate analyzes with distinct SNIa datasets.

    First, we use the Pantheon+ compilation which integrates 20 distinct Type Ia supernova datasets, collectively spanning the redshift interval $0.00122 \leq z \leq 2.26137$ \cite{Scolnic_2022}. It comprises 1701 light curves associated with 1550 spectroscopically confirmed SNe~Ia. 
    
    We also employ the Union3 compilation \cite{Rubin:2023ovl} the most up-to-date version of the Union series, consisting of 2087 supernovae grouped into 22 redshift points covering the range $0.05 \leq z \leq 2.26$. 

    Finally, we add the DES-Dovekie likelihood \cite{Popovic:2025cos,Popovic:2025dov}, which comes as a recalibrated analysis of the Dark Energy Survey 5-year (DESY5) programme \cite{DES:2024tys}. The DESY5 data set encompasses 1635 supernovae over the range $0.0596 \leq z \leq 1.12$, supplemented by 194 high-quality external SNe~Ia at $z < 0.1$, yielding a total of 1829 objects. The DES-Dovekie calibration addresses an incomplete propagation of photometric calibration uncertainties and a numerical approximation in the host-galaxy color law.

\end{itemize}

\subsection{Results}\label{sec:4}

\begin{figure}
    \centering
    \includegraphics[width=\columnwidth]{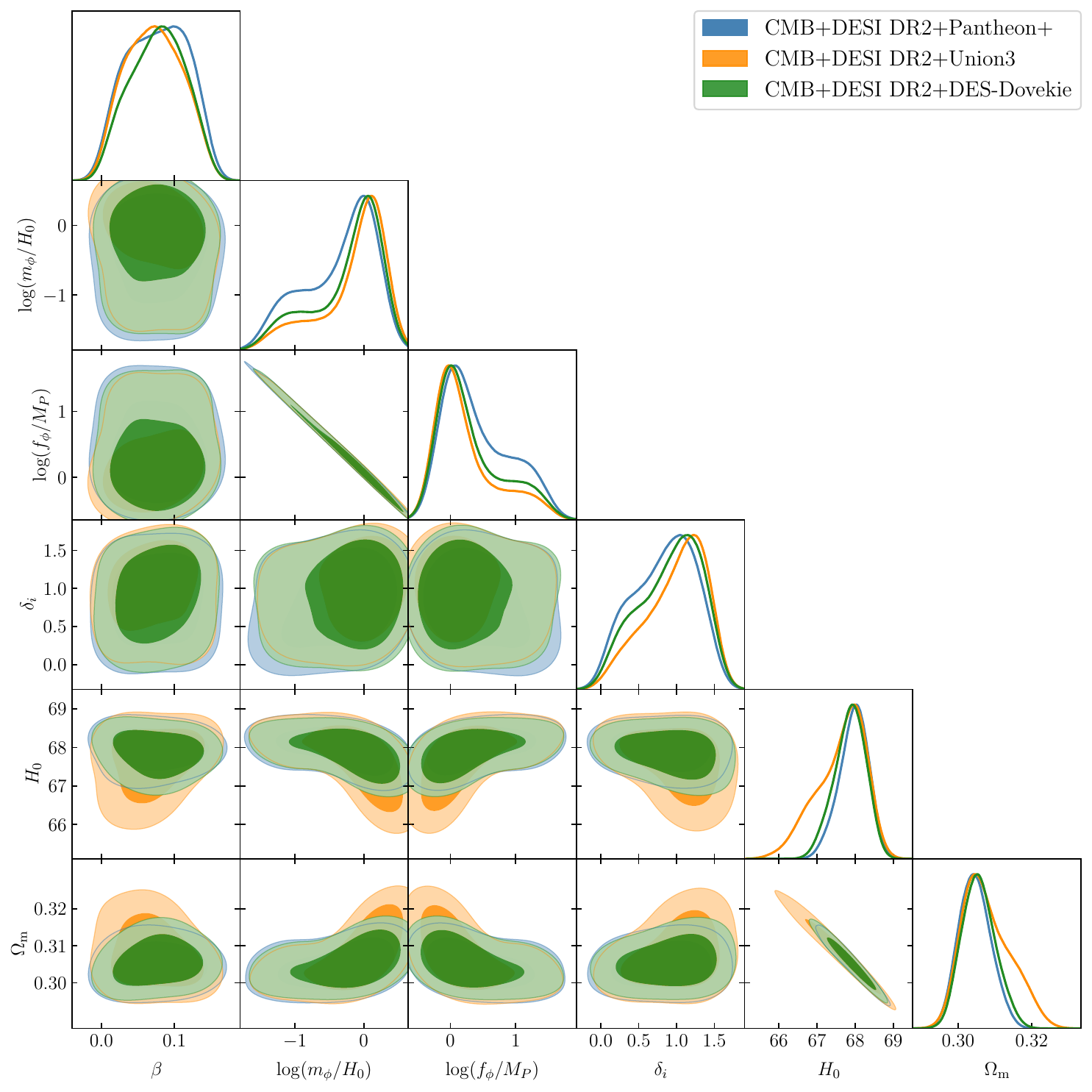}
    \caption{Triangle plots for the interacting axion model with $\delta_i$ as a free parameter. The baseline (CMB+DESI DR2) datasets are combined with the Pantheon+, Union3 and DES-Dovekie SNIa samples in blue, orange and green, respectively.}
    \label{fig: triangle plot full}
\end{figure}

\begin{table*}[ht]
  \centering
  \renewcommand{\arraystretch}{1.4}
  \begin{tabular}{lccc}
  \hline\hline
  Parameter & CMB+DESI DR2+Pantheon\texttt{+} & CMB+DESI DR2+Union3 & CMB+DESI DR2+DES-Dovekie \\
  \hline
  $\beta$ & $0.077^{+0.047}_{-0.042}\,(0.108)$ & $0.072^{+0.039}_{-0.043}\,(0.068)$ & $0.077^{+0.043}_{-0.037}\,(0.082)$ \\
  $\log(m_\phi/H_0)$ & $-0.346^{+0.681}_{-0.342}\,(0.058)$ & $-0.146^{+0.584}_{-0.189}\,(0.274)$ & $-0.226^{+0.632}_{-0.232}\,(0.324)$ \\
  $\log(f_\phi/M_P)$ & $0.406^{+0.346}_{-0.673}\,(0.056)$ & $0.224^{+0.196}_{-0.571}\,(-0.140)$ & $0.294^{+0.241}_{-0.627}\,(-0.274)$ \\
  $\delta_i$ & $0.839^{+0.519}_{-0.359}\,(1.383)$ & $0.997^{+0.504}_{-0.275}\,(1.403)$ & $0.918^{+0.535}_{-0.318}\,(1.008)$ \\
  $H_0$ & $67.96^{+0.43}_{-0.34}\,(67.56)$ & $67.64^{+0.80}_{-0.46}\,(66.14)$ & $67.86^{+0.47}_{-0.37}\,(67.75)$ \\
  $\Omega_{\rm m}$ & $0.3046^{+0.0040}_{-0.0048}\,(0.3085)$ & $0.3077^{+0.0050}_{-0.0080}\,(0.3213)$ & $0.3056^{+0.0042}_{-0.0050}\,(0.3065)$ \\
  \hline
  $\Delta\chi^2$ & $-8.58$ & $-13.68$ & $-13.37$ \\
  \hline
  $\ln B$ & $-0.44 \pm 0.64$ & $+0.25 \pm 0.64$ & $+0.95 \pm 0.63$ \\
  \hline\hline
  \end{tabular}
  \caption{68\% credible intervals for the interacting axion model. Values in parentheses give the best-fit (maximum-likelihood) point for each dataset combination. The second-to-last row gives the $\Delta\chi^2 \equiv \chi^2_\mathrm{axion} - \chi^2_{\Lambda\mathrm{CDM}}$ statistic. The final row gives the Bayes factor $\ln B = \ln\mathcal{Z}_\mathrm{axion} - \ln\mathcal{Z}_{\Lambda\mathrm{CDM}}$, where positive values favor the axion model.}
  \label{tab:constraints_full}
\end{table*}

\begin{figure*}
    \centering
    \includegraphics[width=\linewidth]{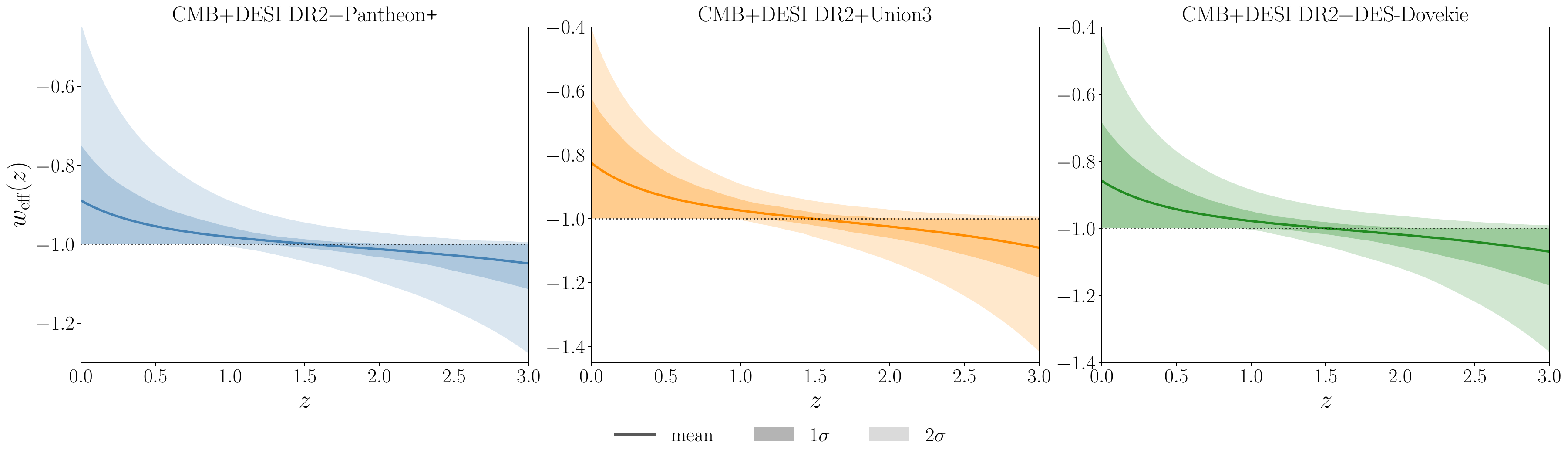}
    \caption{The MC reconstruction of $w_\mathrm{eff}(z)$ as per Eq.~\eqref{w_eff}, based on the statistical distribution of sampled parameters for the CMB+DESI DR2+SNIa data combination, with Pantheon+ (blue), Union3 (orange) and DES-Dovekie (green). The shaded regions depict the 1 and 2 $\sigma$ confidence interval.}
    \label{fig: w_eff_all}
\end{figure*}

Our main results are shown in Fig. \ref{fig: triangle plot full} and Table \ref{tab:constraints_full}, where we show the contour plots for the sampled parameters and their 68\% confidence level distribution, respectively.

As seen in the triangle plots, Fig. \ref{fig: triangle plot full}, the $\log(f_\phi/M_P)$ - $\beta$ contour occupies some of the negative $\log(f_\phi/M_P)$ region for $\beta > 0$, highlighting an available parameter space with a sub-Planckian decay constant. This can be understood in terms of the correlation between the DE axion mass $m_\phi$ and $f_\phi$. Given that $\Lambda_\phi^4 = m_\phi^2 f_\phi^2$ is driven towards $\sim H_0^2 M_P^2$ in order to achieve the required present-day dark energy density, $m_\phi$ becomes largely anti-correlated with $f_\phi$, as seen by the inclination in the $\log(m_\phi/H_0)$ - $\log(f_\phi/M_P)$ contour plot. Therefore, since the posterior of $\log(m_\phi/H_0)$ peaks around $\gtrsim 0$, the posterior peak of $\log(f_\phi/M_P)$ is driven towards the $\lesssim 0$, even though its statistical distribution extends into the positive region.

A marginal sub-Planckian parameter space for $f_\phi$ can also be obtained in standard uncoupled axion DE scenarios \cite{Lin:2025gne}, however, the key difference here is that this is achieved with the initial field value being close to the effective potential minimum, as seen by the imposed prior on $\delta_i$, completely avoiding fine-tuning of the initial conditions. In fact, the best-fit value for $\log(f_\phi/M_P)$ -- defined as the point in parameter space that maximizes the combined likelihood -- is negative for two of the three dataset combinations used, namely with the Union3 and DES-Dovekie SNIa samples. The DES-Dovekie likelihood yields the smallest $f_\phi$, which peaks around $f_\phi \sim 0.53~M_p$.

Regarding the comparison with $\Lambda$CDM, the two-axion model consistently yields a better fit to the data -- represented by a maximum-likelihood difference of $\Delta \chi^2 \sim -13$ with the Union3 and DES-Dovekie SNIa samples -- which is expected in light of the three extra parameters. In order to take into account the extra number of degrees of freedom, as well as the chosen prior distribution, we perform a statistical comparison based on the Bayes factor, defined by
\begin{equation}
    \ln B_{1/2} \equiv \ln \mathcal{Z}_1 - \ln\mathcal{Z}_2\;,
\end{equation}
where -- given a parameter space $\Theta$ and a data set $d$ -- $\mathcal{Z}_i$ is the evidence associated with model $i$, obtained by the integral sum of the likelihood $\mathcal{L}_i(d|\Theta)$ under the prior $\pi_i(\Theta)$, namely $\mathcal{Z}_i = \int d\Theta\mathcal{L}_i(d|\Theta)\pi_i(\Theta)$ \cite{Trotta:2017wnx,Trotta:2008qt}. We read off each of the models' evidences yielded by \texttt{Polychord} with the \texttt{anesthetic} \cite{anesthetic} package and, since the analyses are independent, obtain the error in the Bayes factor by quadrature-summing each evidence individual error, namely $\sigma(\ln B) = \sqrt{\sigma(\ln \mathcal{Z}_1)^2 + \sigma(\ln \mathcal{Z}_2)^2}$. The results are quoted in the final row of Table \ref{tab:constraints_full}.

According to the Jeffreys' scale \cite{jeffreys1998theory}, all of the datasets yield an inconclusive evidence either for or against the two-axion model. Since this model has three extra parameters in comparison to $\Lambda$CDM, one would expect a decrease of its evidence coming from the Occam razor's effect, which penalizes models that introduce extra parameters that are highly constrained by the data \cite{Trotta:2017wnx}. Given that this is not the case here, and by inspecting Fig. \ref{fig: triangle plot full}, one can see that there is a large available parameter space for the three extra parameters $\beta$, $\log(m_\phi/H_0)$ and $\delta_i$. Therefore, even though the constraints are still not tight enough to imply a clear preference for the coupled axion case, the modest prior-to-posterior reduction makes the model competitive with $\Lambda$CDM from the Bayesian point of view. 

Finally, we check how the effective DE equation of state of the interacting axion model preferred by data compares to the CPL EoS implied by DESI. To this end, we compute $w_\mathrm{eff}(z)$, {\color{blue} Eq.~\eqref{w_eff}}, with \texttt{CLASS}, which becomes a function of the random variables $\beta, \log(m_\phi/H_0), H_0, \omega_b, \Omega_c$ and $\delta_i$. We then draw $N = 500$ samples from the combined posterior distribution of these sampled parameters in order to make a Monte Carlo (MC) reconstruction of the effective DE EoS, which is obtained over the redshift range $z \in [0,3]$. The results are shown in Fig. \ref{fig: w_eff_all}. As seen in the plot, the interacting axion model provides an effective DE EoS preferred by data that also crosses to the phantom regime at an earlier time than for the CPL parameterization, centered around $z \sim 1.5$. However, when the uncertainties are taken into account, the reconstructions overlap inside the $\sim$ 2 $\sigma$ bands. 

\section{Discussion and Conclusions}\label{sec:5}
The recent findings from the DESI Collaboration \cite{DESI:2025zgx} have hinted at the possibility that the dark sector might feature richer dynamics than what is predicted by standard $\Lambda$CDM cosmology. In general, an interacting dark sector is capable of producing an expansion rate that yields an effective DE EoS which crosses the phantom divide and is consistent with that preferred by the DESI data.

In this work, we have investigated this scenario in the context of two interacting axion-like fields motivated by the fact that such interactions can arise in string compactification models and axion monodromies. In light of the necessary mass hierarchies that these fields have to obey in order to behave like DE ($\phi$ field) and DM ($\chi$ field), we were able to effectively describe the dynamics of the heavier DM axion in terms of a quadratic potential with a DE-dependent mass term. Given the lightness of the DE axion -- which results in a slow-roll evolution across most of the cosmic expansion -- the DM axion mass evolves slowly compared to its oscillation period, which allows us to express its dynamics in the coupled quintessence framework, where it behaves like a pressureless fluid.  Since the initial energy density in DM is much larger than DE, we have seen, Eq.~\eqref{rho_scaling_RD_and_MD}, that the energy density of the DM fluid is quickly driven to an attractor that scales as $a^{-2}$ during radiation domination -- where the source term $Q(a)$, Eq.~\eqref{Q and gamma of phi explict}, is negative -- and $a^{-3}$ during matter domination, before staying roughly constant at present times, when $Q$ changes to positive values. This change of behavior in the coupling function -- which follows directly from the DE scalar field dynamics and its effective potential transition -- is key to producing the late-time observables consistent with DESI's findings.

We then performed a statistical analysis with current data to check on the observational viability of the model. We found that it generally yields a better fit than $\Lambda$CDM with the datasets employed, which is expected given the higher dimensionality of the parameter space. However, the two-axion scenario is not over-penalized by the extra number of free parameters, as seen by the inconclusive evidence in the Bayes ratio test, staying competitive with $\Lambda$CDM. 

Therefore, the coupled quintessence scenario is still competitive as an interpretation of current data and can be accommodated in the context of axion phenomenology, which can be embedded into several theoretical constructions, which is very promising. We hope that future precision data may help to further distinguish these models and shed additional light into the nature of the cosmological dark sector.

\section*{Acknowledgments}
We would like to thank Justin Khoury for helpful discussions. RdS thanks the Particle Cosmology group at the University of Nottingham for the incredible hospitality during the period where this work was being developed. RdS is supported by the Coordena\c{c}\~ao de Aperfei\c{c}oamento de Pessoal de N\'ivel Superior (CAPES) and this work was developed under the support of the Programa de Doutorado Sandu\'iche no Exterior (PDSE) Fellowship. EJC is supported by an STFC Consolidated Grant [Grant No. ST/X000672/1]. JA is supported by Conselho Nacional de Desenvolvimento Científico e Tecnológico (CNPq) grant No. 307683/2022-2 and Funda\c{c}\~ao de Amparo \`a Pesquisa do Estado do Rio de Janeiro (FAPERJ) grant No. 299312 (2023). The development of this work was aided by the National Observatory Data Center (CPDON).

\appendix
\section{Numerical Solutions and WKB Approximation}\label{app:A}

\begin{figure*}
    \centering
    \includegraphics[width=\linewidth]{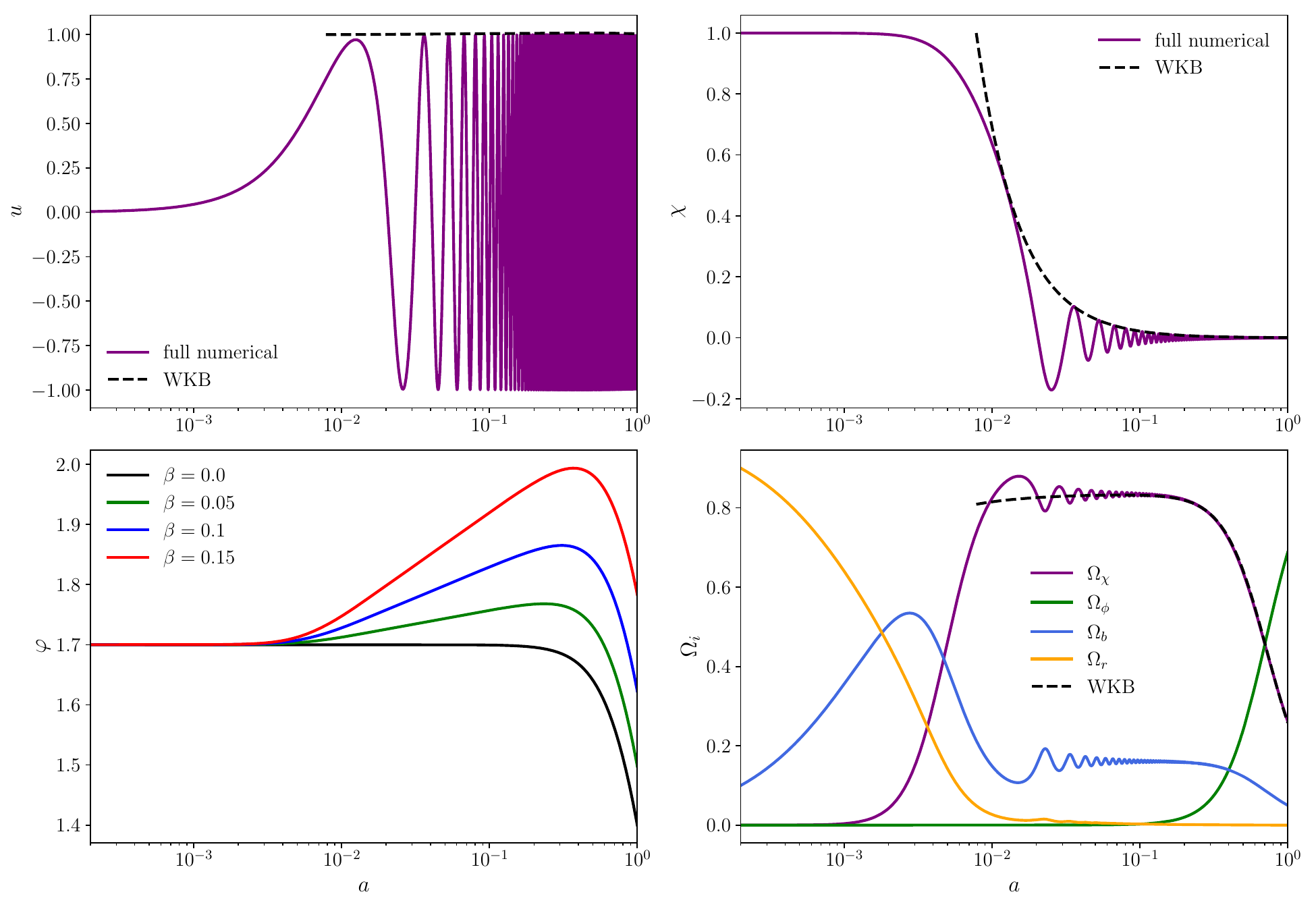}
    \caption{Full numerical evolution of $u(t)$ (upper left) and $\chi(t)$ (upper right), with the WKB approximation for the latter represented by the dashed curve. In the bottom left, we show the numerical evolution of $\varphi$ for increasing coupling strength and the behavior of the density parameters in the bottom right plot, with the WKB approximation also in dashed.}
    \label{fig: two-axion-background}
\end{figure*}

As discussed in the text, expressing the two-axion system as an effective coupled quintessence framework depends on the validity of the WKB approximation to the DM axion dynamics. Below, we describe why we believe this is the case, by considering a model where we can follow the full numerical evolution and compare directly with the WKB approximation. The price we have to pay, is that we introduce an unphysically small DM mass in the numerical case, $m_\chi = 6.7\cdot 10^{2} ~m_\phi$, but we believe the principle holds for the more realistic case too. In its full generality, the equation of motion, Eq.~\eqref{u EOM}, for the rescaled variable $u(t)$ written in terms of the number of e-folds $N = \ln a$ is given by
\begin{equation}\label{u EOM in terms of N}
    u'' + \frac{E'}{E} u ' + \mu_u^2(N) u = 0\;,
\end{equation}
where, for this Appendix only primes now denote derivatives with respect to $N$, $E \equiv H/H_0$ and $\mu_u^2(N) = \frac{m_\chi^2(\varphi)}{H_0^2 E^2} - \frac{9}{4} - \frac{3}{2}\frac{E'}{E}$ is the dimensionless time-dependent mass.

The KG equation Eq.~\eqref{KG eq} for $\varphi$ reads
\begin{equation}\label{varphi EOM wrt N}
    \varphi'' + \left(\frac{E'}{E} + 3\right)\varphi' + \left(\frac{m_\phi^2}{H_0^2 E^2} - \frac{m_\chi^4 a^{-3}}{2f_\phi^2 H_0^2 E^2}\beta u^2\right)\sin\varphi = 0\;.
\end{equation}

Eqs. \eqref{u EOM in terms of N} and \eqref{varphi EOM wrt N} are supplemented by the dimensionless quadratic Hubble rate and its normalized derivative, namely:

\begin{widetext}
\begin{equation}\label{sound horizon CMB}
E^2 = \frac{\Omega_b(N) + \Omega_r(N) + \frac{m_\chi^4 a^{-3}}{6 M_P^2 H_0^2}(1+\beta\cos\varphi)u^2 + \frac{m_\phi^2 f_\phi^2}{3 M_P^2 H_0^2}(1-\cos\varphi)}{1 - \frac{1}{6}\left[ \frac{m_\chi^2 a^{-3}}{M_P^2}(9 u^2/4 - 3 u u' + {u'}^2) + \frac{f_\phi^2}{M_P^2}{\varphi'}^2 \right]}\end{equation}
\end{widetext}
and
\begin{equation}\label{E_prime_E}
    \begin{split}
        \frac{E'}{E}  = &-\frac{3 \Omega_b(N) + 4\Omega_r(N)}{2 E^2} \\ &- \frac{1}{2}\left[\frac{m_\chi^2 a^{-3}}{M_P^2}(9 u^2/4 - 3 u u' + {u'}^2) + \frac{f_\phi^2}{M_P^2}{\varphi'}^2\right]\;.
    \end{split}
\end{equation}

First, we numerically solve Eqs. \eqref{u EOM in terms of N} - \eqref{E_prime_E} in their full generality for the variables $u, u', \varphi, \varphi'$, with no approximations taken. To test the validity of the WKB approximation, we also solve the system assuming Eq. \eqref{u(t) WKB}, where the envelope dynamics of $u(t)$ is given in terms of the amplitude $A(t) = C/\sqrt{m_u(t)}$. Hence, we solve for $\varphi,\varphi',A$ according to Eq.~\eqref{varphi EOM wrt N} along with the constraint equation for $A'$
\begin{equation}
    \frac{A'}{A} = -\frac{1}{2}\frac{m_u'}{m_u} \sim -\frac{1}{2}\frac{m_\chi'(\varphi)}{m_\chi(\varphi)} = \frac{\beta\sin\varphi}{4(1+\beta\cos\varphi)}\varphi'\;,
\end{equation}
where we have approximated $m_u(t) \sim m_\chi(\varphi)$ in Eq.~\eqref{msqu} since $m_\chi \gg H$ deep in the oscillatory phase of $\chi$, and finally used Eq.~\eqref{V eff DM and m_chi(phi)} for $m_\chi(\varphi)$. 

We show in Fig. \ref{fig: two-axion-background} the solutions for the main background quantities for $m_\phi = 1.5 H_0$, $\delta_i = 1.7$ and $\beta = 0.1$, which are the same reference values used in the coupled quintessence discussion of Sec. \ref{sec: Background evolution and apparent phantom-crossing}. For the axion DM mass, we use $m_\chi = 6.7\cdot 10^{2} ~m_\phi$, in order to obtain a sensible cosmology with a current DM density of $\Omega_{\chi,0} \sim 0.26$. Again, we shoot on $f_\phi$ to satisfy the closure condition and for this parameter combination we obtain $f_\phi \sim 0.9 M_P$. For the full solution, we start the evolution shortly before the transition to matter domination, at $z_i = 5000$ with $\chi_i = 1.0$ and $\chi'_i = 0$. In the case of the WKB approximation, we start evolving the system when the field $u$ begins oscillating around its minimum, which happens deep into the matter-dominated epoch at $z_i \sim 100$.

We stress that the choice of $m_\chi$ is not unique and different values with other combinations of $m_\phi$, $\delta_i$ and $\beta$ yield similar dynamics. However, once $m_\chi \gtrsim 10^3~m_\phi$, the system becomes significantly stiff as the integrator of the $u$ variable passes through zero many times, making these simulations computationally expensive. Nevertheless, it does not affect our purposes here since our goal is only to demonstrate the validity of the WKB approximation once $u$ enters the oscillatory regime. 

The upper left panel {\color{blue} of Fig. \ref{fig: two-axion-background}} shows the evolution of $u(t)$ while in the upper right we have $\chi(t) = a^{-3/2} u(t)$, with the black dashed curve displaying the WKB approximation. It is clear that the amplitude of $u$ stays approximately constant, which means that the full solution for $\chi$ does evolve in the WKB branch with its envelope scaling as $a^{-3/2}$. In the bottom left panel we also show the evolution of $\varphi$ from Eq. \eqref{varphi EOM wrt N}, which reproduces the overall behavior seen in the left panel of Fig. \ref{fig: varphi, varphi_prime and V_eff}, where the fluid approximation for $\chi$ in the coupled quintessence framework was taken (Eq. \eqref{rho_chi scaling}). Finally, the bottom right panel displays the evolution of the energy density parameters $\Omega_i \equiv \frac{\rho_i}{3 M_P^2 H^2}$ for different species, including the standard baryonic and radiation fluids. The WKB approximation for $\Omega_\chi$ also captures the overall evolution of the full oscillating solution.

For the order of magnitude of $m_\chi \sim 10^{3}~m_\phi$ taken in the full numerical evaluation of the fields' background dynamics -- Eqs \eqref{u EOM in terms of N} and \eqref{varphi EOM wrt N} -- the DM axion stays frozen at early times and starts behaving like a pressureless fluid at $z_i \sim 100$, when the oscillatory motion begins. Therefore, taking the present-day density to be $\Omega_{\chi,0} \sim 0.26$, its energy density before the onset of the oscillation period is lower than that expected from the $a^{-3}$ scaling, resulting in the intermediate baryonic domination observed in Fig. \ref{fig: two-axion-background}. This is simply a consequence of not taking the DM axion to be heavy enough, and full CDM domination is reestablished once the field begins oscillating before the matter-radiation transition epoch, which always happens for the DM mass range considered in the main text analyses -- see Sec. \ref{sec:1}.

\begin{figure}
    \centering
    \includegraphics[width=\linewidth]{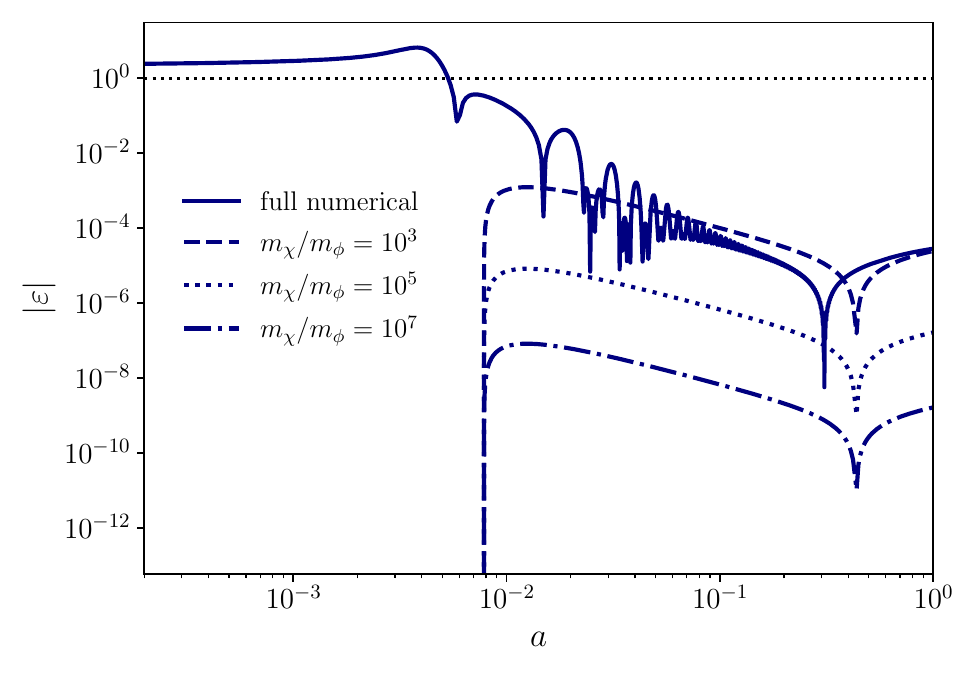}
    \caption{Evolution of the WKB parameter $\varepsilon \equiv \dot m_u/m_u^2$ for the full numerical solution of the background two-axion system (solid purple) assuming Eq.~\eqref{epsilon explicit}, as well as for the WKB approximation (dashed). In the WKB approach, we also show the behavior for increasing DM masses.}
    \label{fig: two-axion-epsilon}
\end{figure}

Finally, the WKB approach can be summarized by checking the smallness of the parameter $\varepsilon \equiv \dot m_u/m_u^2$ -- introduced in the discussion around Eq. \eqref{u(t) WKB}. Note that $\varepsilon = \frac{H}{m_u}\frac{m_u '}{m_u} \sim \frac{H}{m_\chi(\varphi)}\frac{m_\chi'(\varphi)}{m_\chi(\varphi)}$, which will be suppressed by our initial assumption that $m_\chi \gg H$. In terms of $m_u^2(t) = m_\chi^2(\varphi(t)) - \frac{9}{4}H^2 - \frac{3}{2}\dot H$, $\varepsilon$ can be written explicitly as
\begin{equation}\label{epsilon explicit}
    \varepsilon = \frac{\frac{-\beta \sin\varphi}{2(1+\beta\cos\varphi)}\varphi' - \frac{3\left[3E'/E - (E'/E)^2 - E''/E\right]}{4\left[\frac{m_\chi^2}{H^2}(1+\beta \cos\varphi) - 9/4 - (3/2)E'/E\right]}}{\sqrt{\frac{m_\chi^2}{H^2}(1+\beta \cos\varphi) - 9/4 - (3/2)E'/E}}\;,
\end{equation}
which reduces to $\varepsilon\sim \frac{H}{m_\chi(\varphi)}\frac{m_\chi'(\varphi)}{m_\chi(\varphi)}$ in the $m_\chi \gg H$ limit.

In the left panel of Fig. \ref{fig: two-axion-epsilon}, we show the behavior of the absolute value of $\varepsilon$ for both the full numerical solution -- per Eq.~\eqref{epsilon explicit} -- in solid blue and also assuming the WKB regime, taking $\varepsilon\sim \frac{H}{m_\chi(\varphi)}\frac{m_\chi'(\varphi)}{m_\chi(\varphi)}$, as dashed/dotted curves. For these latter curves, we consider three representative cases of increasing DM mass. Thus, while $\varepsilon$ is large early on in the full numerical approach, it quickly drops well below unity as the field goes deeper into the oscillatory regime, where it is well approximated by the WKB solution. Moreover, considering heavier DM masses (which cannot be solved in full generality due to stiffness in the EOM) further improves the approximation, meaning that the field can safely be described by an effective pressureless fluid in these mass ranges.

\bibliography{references}

\end{document}